\documentclass{iopconfser}
\usepackage{amssymb}
\usepackage{amsmath}
\usepackage{amsfonts}
\usepackage{graphicx}
\usepackage{hyperref}

\usepackage{orcidlink}
\begin{document}

\title{Instability mechanisms on asymptotically AdS black hole backgrounds}

\author{Filip Ficek$^{1,2}$\orcidlink{0000-0001-5885-7064} and Maciej Maliborski$^{1,2,3}$\orcidlink{0000-0002-8621-9761}}

\affil{$^1$University of Vienna, Faculty of Mathematics, Oskar-Morgenstern-Platz 1, 1090 Vienna, Austria}
\affil{$^2$University of Vienna, Gravitational Physics, Boltzmanngasse 5, 1090 Vienna, Austria}
\affil{$^3$TU Wien, Institute of Analysis and Scientific Computing, Wiedner Hauptstraße 8–10, 1040 Vienna, Austria}

\email{filip.ficek@univie.ac.at, maciej.maliborski@univie.ac.at}

\begin{abstract}
We investigate dynamics of the conformal scalar equation with defocusing cubic nonlinearity and Robin boundary conditions on the Reissner-Nordstr\"{o}m-anti-de Sitter background. At certain parts of the parameters phase space the zero solution ceases to be stable. We present two mechanisms responsible for this phenomenon.
\end{abstract}

\section{Introduction}
Dynamics of a scalar field in asymptotically AdS spacetimes has been studied in the past in various contexts \cite{Hor00, Mos02, Car03, Win03, Ber09,Kon11,War15,Har23}. This note belongs to the line of works investigating existence and stability of static solutions in such set-ups \cite{Biz20,FFMM24,FFMM25}. In particular, in \cite{FFMM25} we demonstrated how the zero field loses stability for certain choices of the boundary condition and parameters of the background Reissner-Nordstr\"{o}m-anti-de Sitter (RNAdS) metric. There, the analysis was conducted from the point of view of the bifurcation theory: we studied how the stability of zero changes due to a bifurcation of non-trivial static solutions. The goal of this work is to take another approach, based on the energy, to identify mechanisms responsible for this phenomenon.

\section{Setup}
We consider RNAdS background in standard coordinates. However, instead of the common parametrisation provided by BH mass $M$, its charge $Q$ and the AdS radius $\ell$, it is more convenient to introduce the event horizon radius $r_H$ and the charge parameter $\sigma=r_C/r_H$, where $r_C$ is the Cauchy horizon radius (so $\sigma=0$ corresponds to the Schwarzschild-AdS spacetime and $\sigma=1$ gives an extremal RNAdS). If we additionally set the scale by fixing $\ell=1$, the line element takes the form of
\begin{equation}
    \mbox{d}s^2=-V(r) \mbox{d}t^2+V(r)^{-1}\mbox{d}r^2+r^2 \mbox{d}\Omega^2,
\end{equation}
where
\begin{equation}\label{eq:V}
    V(r)=1-\frac{(1+\sigma)r_H}{r}\left[1+(1+\sigma^2)r_H^2\right]+\frac{\sigma\,r_H^2}{r^2}\left[1+(1+\sigma+\sigma^2)r_H^2\right]+r^2.
\end{equation}
For the convenience we replace time $t$ with the null coordinate $v$ defined by $\mbox{d}v=\mbox{d}t+\mbox{d}r/V$. We also perform the compactification of the space by introducing $y=1/r$ and $y_H=1/r_H$. By slightly abusing the notation, it lets us consider the function $V(y)$ that is used in the remainder.

We study a conformally coupled defocusing cubic scalar field
\begin{equation}\label{eq:field}
    \Box\phi+2\phi-\phi^3=0,
\end{equation}
where $\Box$ is the wave operator on the background given by the introduced metric. For the further convenience we limit ourselves to spherically-symmetric solutions $\phi=\phi(v,y)$. We also redefine the dependent variable $\Phi(v,y)=\phi(v,y)/y$ so Eq.~\eqref{eq:field} eventually becomes
\begin{equation}\label{eq:field_2}
    2\partial_y\partial_v\Phi-\partial_y\left(y^2\, V(y)\, \partial_y\Phi\right)-\left(y\, V'(y)+\frac{2}{y^2}\right)\Phi+\Phi^3=0.
\end{equation}


Due to the presence of the conformal term, this equation can be equipped with the Robin boundary condition (BC) at infinity ($y=0$), giving rise to a one-dimensional family of the problems\footnote{It stays in contrast to the case of a scalar field with minimal coupling, where only Dirichlet boundary condition (BC) leads to the square integrability of the field \cite{Biz14, FFMM24}.}. We parametrise them by the angle $\beta\in(-\pi/2,\pi/2]$ defined by
\begin{equation}\label{eq:BC}
    \left.\left[(-\partial_v\Phi+\partial_y\Phi)\cos\beta -\Phi\sin\beta\right]\right|_{y=0}=0.
\end{equation}
Let us point out, that fixing $\beta=0$ leads to Neumann BC, while $\beta=\pi/2$ recovers Dirichlet BC.


\section{Dynamics}
\begin{figure}[t!]
    \centering
    \includegraphics[width=0.32\linewidth]{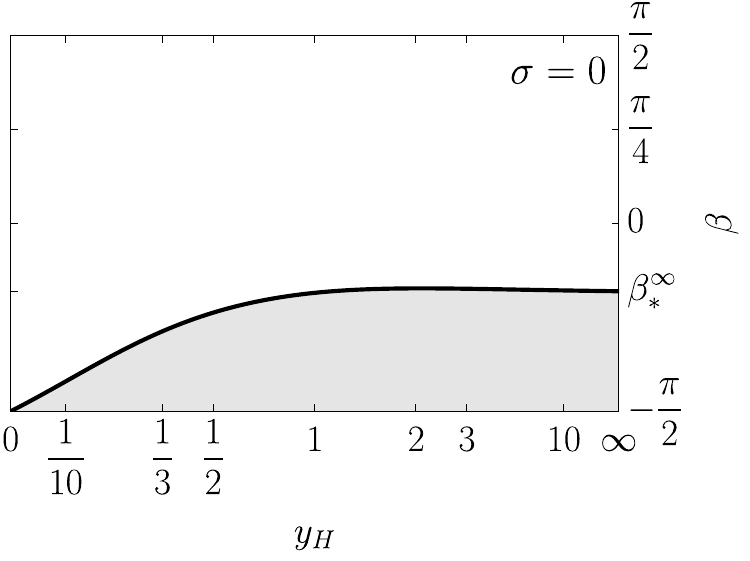}
    \includegraphics[width=0.32\linewidth]{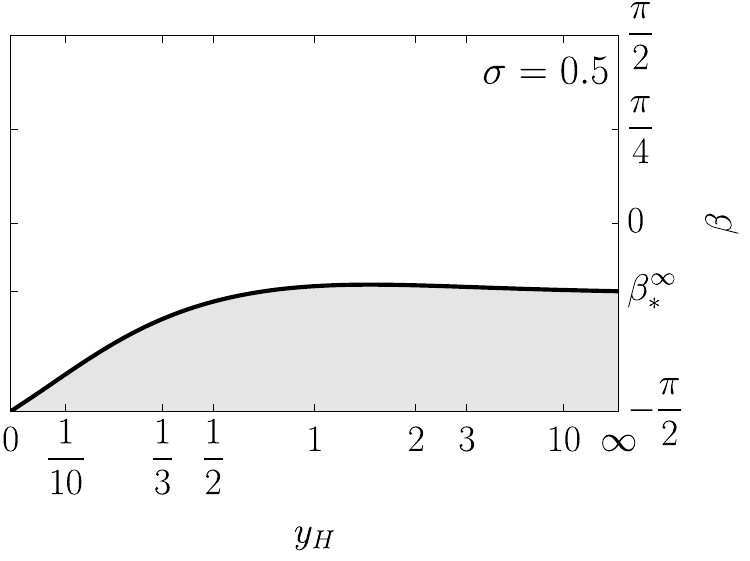}
    \includegraphics[width=0.32\linewidth]{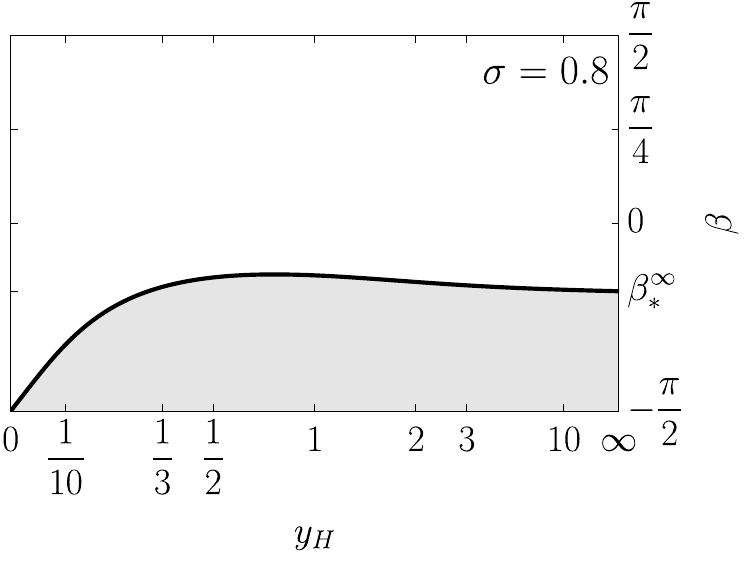}\\
    \includegraphics[width=0.32\linewidth]{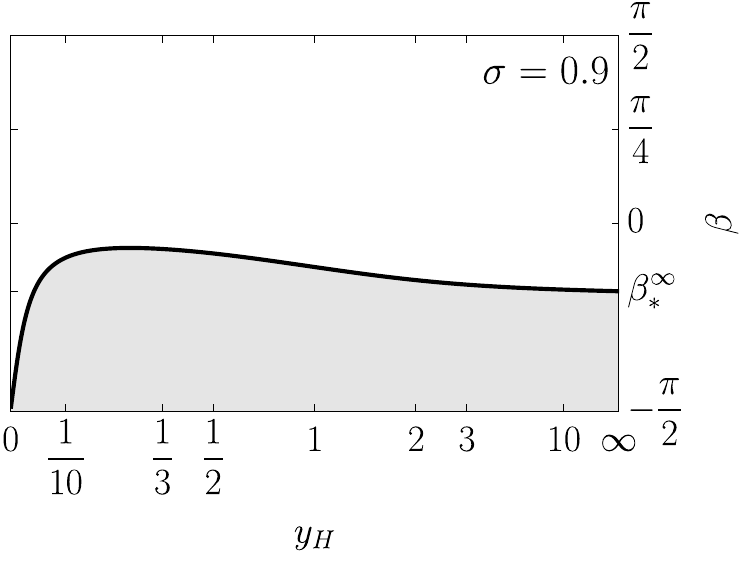}
    \includegraphics[width=0.32\linewidth]{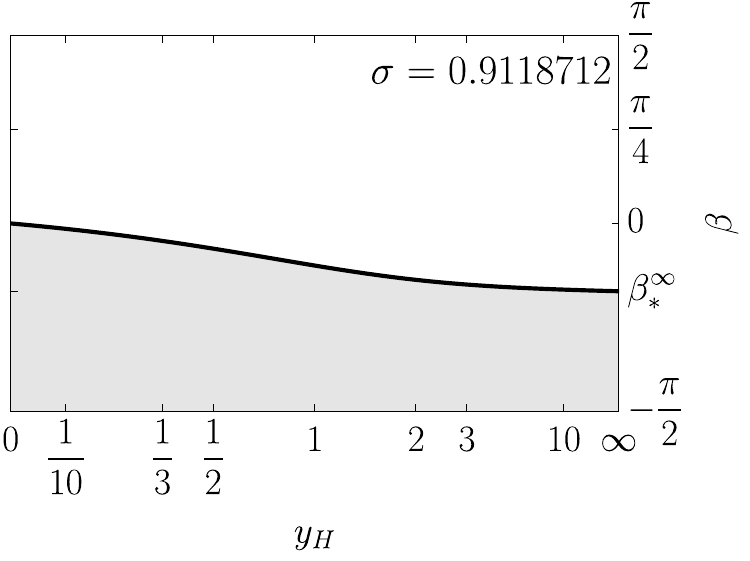}
    \includegraphics[width=0.32\linewidth]{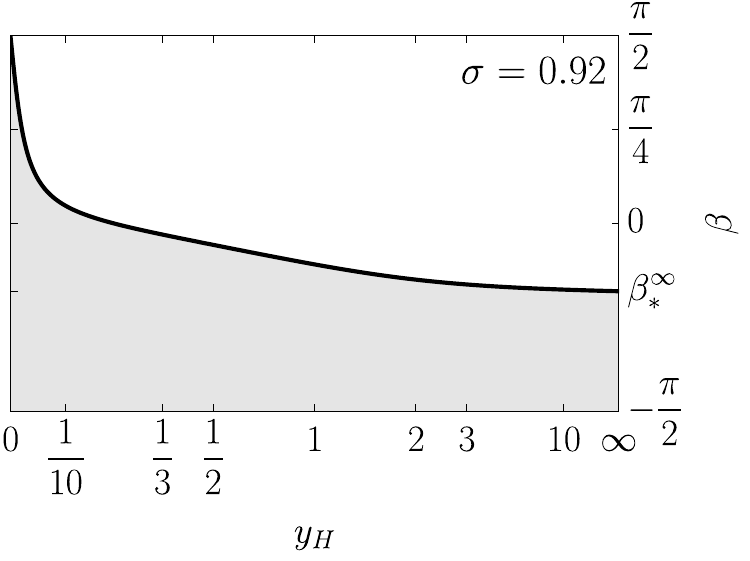}\\   
    \includegraphics[width=0.32\linewidth]{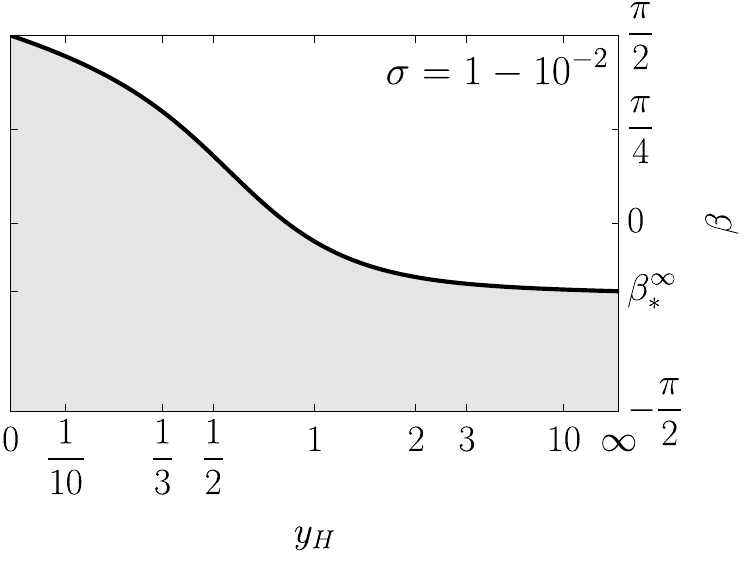}
    \includegraphics[width=0.32\linewidth]{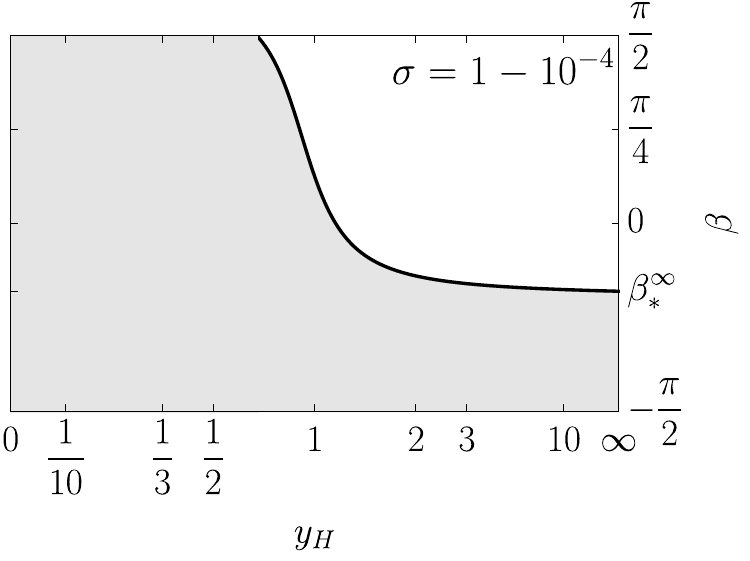}
    \includegraphics[width=0.32\linewidth]{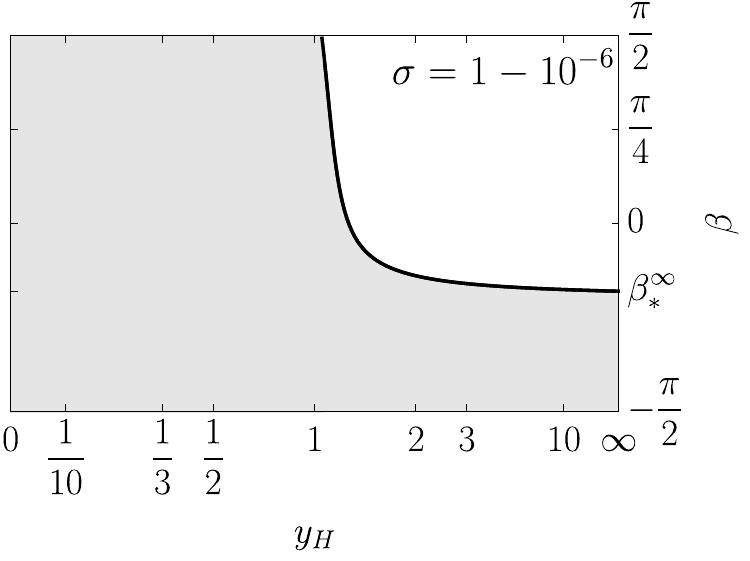}\\ 
    \caption{Stability phase plots for various values of $\sigma$. White colour indicates regions where the zero solution is stable, while in gray areas it is unstable.}
    \label{fig:phase_plots}
\end{figure}

Numerical simulations show a dichotomy in the evolution of the field described by Eq.~\eqref{eq:field_2}. One possible behaviour is a decay of the field to zero. Alternatively, the field may converge to a certain static solution $s=s(y)$ or its negative copy $-s$. Interestingly, which of these two scenarios plays out depends only on the parameters $y_H$, $\sigma$, and $\beta$, not on the specific initial data (as long as they are non-trivial). Similarly the profiles of the attractors $s$ also depend only on those parameters.
As a result, we can use simple phase plots presented in Fig.~\ref{fig:phase_plots} to describe the dynamics of this system. White parts of these plots indicate parameters values for which field always decays to zero, i.e., zero is a stable solution. In the gray regions zero is unstable and any non-trivial initial data converges eventually to a nodeless non-zero static solution $\pm s$. 
Solid black lines in Fig.~\ref{fig:phase_plots} can be interpreted as points of pitchfork bifurcation. At these points the zero solution loses its stability and gives raise to two stable static solutions $\pm s$. Regardless of the charge $\sigma$, in the limit $y_H\to \infty$ (small black holes) the boundary between stable and unstable regions is localised at Robin parameter $\beta_*^\infty=-\arctan(2/\pi)$, cf.\ \cite{Biz20}.

Now let us focus on the large black holes limit ($y_H\to 0$). In the case of SAdS spacetime ($\sigma=0$) the zero solution is stable. This conclusion can be made by observing in Fig.\ \ref{fig:phase_plots} that for any value of $\beta$ one can find sufficiently small $y_H$ so that they belong to the white region of the plot. The same observation holds as the charge increases up to the critical value $\sigma=\sigma_{cr}\approx0.9118712$. Above it, zero becomes unstable in the presence of large black holes: for sufficiently small values of $y_H$ any non-trivial initial data builds up to one of two static solutions $\pm s$.


\section{Instability mechanisms}
To study the nature of the instability of the zero solution, we introduce the following quantity that will be called energy
\begin{equation}\label{eq:energy}
    E=\int_{0}^{y_H}\left(\frac{y^2}{2}V(y)\left(\partial_y\Phi\right)^2-\frac{1}{2}\left(yV'(y)+\frac{2}{y^2}\right)\Phi^2+\frac{1}{4}\Phi^4 \right)\mbox{d}y.
\end{equation}
Then simple calculations using Eqs.~\eqref{eq:field_2} and \eqref{eq:BC} show that $E$ changes with the time coordinate $v$ as 
\begin{equation}\label{eq:flux}
    \frac{\mbox{d}E}{\mbox{d}v}=-\left.\left(\partial_v \Phi\right)^2 \right|_{y=y_H}-\frac{\tan\beta}{2}\left.\partial_v\left(\Phi^2\right)\right|_{y=0}.
\end{equation}
The first term in this expression is non-positive and describes the energy absorption by the black hole. The second term, however, does not have a fixed sign, and it allows for the exchange of energy through Scri. Let us point out that it vanishes identically for both Neumann and Dirichlet BC.

Using the energy we can distinguish two instability mechanisms. The first one concerns the positive definiteness of $E$. Since $V$ is a metric function \eqref{eq:V}, the first term in \eqref{eq:energy} is positive in the considered region of the spacetime ($0<y<y_H$). The same conclusion is trivially true for the third term, leaving just the second term for further analysis. One can show that it is positively defined for any $y_H$ as long as $1-\sigma-\sigma^2-\sigma^3>0$. This condition holds for $\sigma<\sigma_E$ where $\sigma_E\approx 0.5436$. Hence, $\sigma>\sigma_E$ is a necessary condition for existence of static solutions with negative energy. 
In such case it may be possible for a perturbation of the zero solution to lose energy through the boundaries and converge to this static solution. Let us point out that this mechanism has been also recently discussed for a linear case in \cite{Zheng24}.

The fact that the second term in Eq.~\eqref{eq:flux} allows for the influx of energy through the Scri constitutes a second instability mechanism. Indeed, in \cite{War14} it has been shown that the linearised equation \eqref{eq:field_2} with $\sigma=0$ possesses a growing mode for certain values of $y_H$ and $\beta$. As we have demonstrated in \cite{FFMM24} this effect is caused by the energy influx through Scri and is also present for the nonlinear equation. However, according to the observation made above, this instability mechanism shall be absent for Neumann and Dirichlet BCs since there can be no energy influx then.

There are regions of the phase space where only one of these mechanisms is present. For $\beta=0$ or $\beta=\pi/2$ there is no energy transfer through Scri, so any instability comes only from the lack of positive definiteness of the energy. On the other hand, $E$ is positively defined when $\sigma<\sigma_E$, and there the instability has to be caused by the energy influx through Scri. In the remaining parts of the phase space we can observe a complex interplay between these two mechanisms that requires further studies. The main difficulty comes from the fact that Eq.~\eqref{eq:flux} allows for both energy inflow and outflow through Scri. As a result, this energy exchange can serve as both stabilising and destabilising mechanism, depending on the situation. 


\section*{Acknowledgements}
We acknowledge the support of the Austrian Science Fund (FWF) through Project \href{http://doi.org/10.55776/P36455}{P 36455}, Wittgenstein Award \href{http://doi.org/10.55776/Z387}{Z 387}, and the START-Project \href{http://doi.org/10.55776/Y963}{Y 963}.

\bibliography{iopart-num}

\end{document}